\documentclass{ieeeaccess}
\usepackage{cite}
\usepackage{amsmath,amssymb,amsfonts}
\usepackage{algorithmic}
\usepackage{graphicx}
\usepackage{textcomp}
\usepackage{hyperref}    

\usepackage{bm}
\makeatletter
\AtBeginDocument{\DeclareMathVersion{bold}
\SetSymbolFont{operators}{bold}{T1}{times}{b}{n}
\SetSymbolFont{NewLetters}{bold}{T1}{times}{b}{it}
\SetMathAlphabet{\mathrm}{bold}{T1}{times}{b}{n}
\SetMathAlphabet{\mathit}{bold}{T1}{times}{b}{it}
\SetMathAlphabet{\mathbf}{bold}{T1}{times}{b}{n}
\SetMathAlphabet{\mathtt}{bold}{OT1}{pcr}{b}{n}
\SetSymbolFont{symbols}{bold}{OMS}{cmsy}{b}{n}
\renewcommand\boldmath{\@nomath\boldmath\mathversion{bold}}}
\makeatother

\def\BibTeX{{\rm B\kern-.05em{\sc i\kern-.025em b}\kern-.08em
    T\kern-.1667em\lower.7ex\hbox{E}\kern-.125emX}}

\begin{document}
\history{Date of publication xxxx 00, 0000, date of current version xxxx 00, 0000.}
\doi{10.1109/ACCESS.2023.0322000}

\title{MEGA: Maximum-Entropy Genetic Algorithm for Router Nodes Placement in Wireless Mesh Networks}
\author{\uppercase{N. Ussipov}\authorrefmark{1}, 
\uppercase{S. Akhtanov}\authorrefmark{1}, 
\uppercase{D. Turlykozhayeva}\authorrefmark{1},
\uppercase{S. Temesheva}\authorrefmark{1},
\uppercase{A. Akhmetali}\authorrefmark{1},
\uppercase{M. Zaidyn}\authorrefmark{1},   
\uppercase{T. Namazbayev}\authorrefmark{1},
\uppercase{A. Bolysbay}\authorrefmark{1},
\uppercase{A. Akniyazova}\authorrefmark{1},
\uppercase{and Xiao Tang}\authorrefmark{2} , \IEEEmembership{Member, IEEE}}

\address[1]{Department of Solid State Physics and Nonlinear Physics, Al-Farabi Kazakh National University, Almaty, Kazakhstan}
\address[2]{School of Electronics and Information, Northwestern Polytechnical University, Xi’an 710072, China (e-mail: tangxiao@nwpu.edu.cn)}

\tfootnote{This work was supported by the Committee of Science of the Ministry of Science and Higher Education of the Republic of Kazakhstan under Grant AP19674715.}

\markboth
{Ussipov \headeretal: MEGA: Maximum Entropy Genetic Algorithm for mesh router nodes placement }
{Ussipov \headeretal: MEGA: Maximum Entropy Genetic Algorithm for mesh router nodes placement}

\corresp{Corresponding author: Turlykozhayeva D. (e-mail: turlykozhayeva.dana@kaznu.kz).}

\begin{abstract}
Over the past decade, Wireless Mesh Networks (WMNs) have seen significant advancements due to their simple deployment, cost-effectiveness, ease of implementation and reliable service coverage. However, despite these advantages, the placement of nodes in WMNs presents a critical challenge that significantly impacts their performance. This issue is recognized as an NP-hard problem, underscoring the necessity of development optimization algorithms, such as heuristic and metaheuristic approaches. This motivates us to develop the Maximum Entropy Genetic Algorithm (MEGA) to address the issue of mesh router node placement in WMNs. To assess the proposed method, we conducted experiments across various scenarios with different settings, focusing on key metrics such as network connectivity and user coverage.  The simulation results show a comparison of MEGA with other prominent algorithms, such as the Coyote Optimization Algorithm (COA), Firefly Algorithm (FA), Genetic Algorithm (GA), and Particle Swarm Optimization (PSO), revealing MEGA's effectiveness and usability in determining optimal locations for mesh routers.
\end{abstract}


\titlepgskip=-15pt

\maketitle
\section{Introduction}
\label{sec:introduction}
\begin{figure}[h]
    \centering
    \includegraphics[width=1\linewidth]{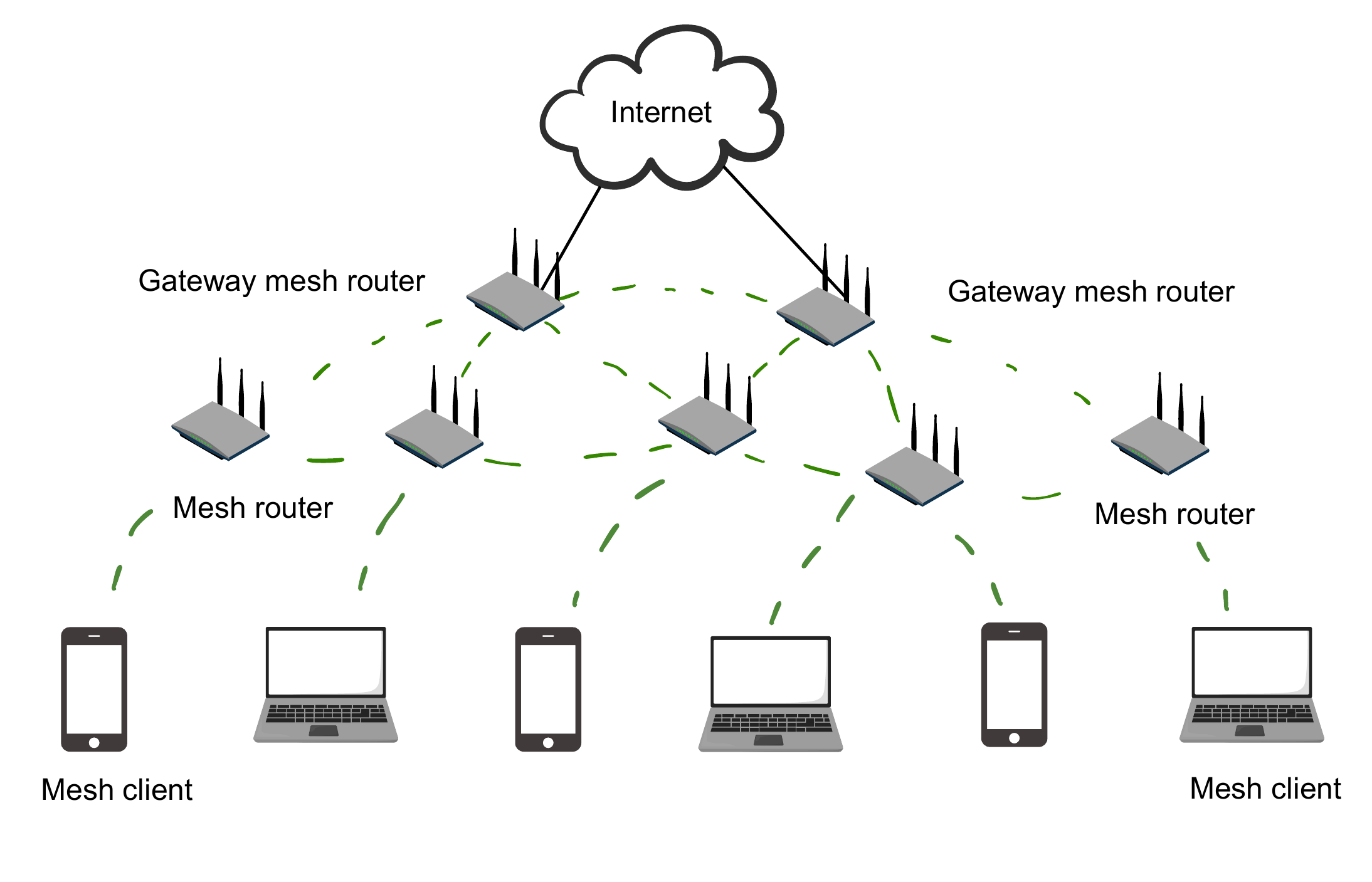}
    \caption{\textbf{Wireless Mesh Network architecture.}}
    \label{figure 1}
\end{figure}

\PARstart {A}s an emerging technology, Wireless Mesh Network (WMN) has gained increasing attention in the communication field during the last decade. This attention is due to the following advantages, such as quick and easy implementation, dynamic self-organization, self-configuration, extensive network coverage and cost effectiveness \cite{Barolli1}, \cite{Hussain2}, \cite{Nouri3}. Also, WMN can support a wide range of applications, e.g., broadband home networking, education field, healthcare, building automation, disaster management, rescue operations, and military \cite{Janjua4}, \cite{Rethfeldt5}.  WMN is made up of three different types of nodes: Mesh Routers (MRs), Mesh Gateways (MGs), and Mesh Clients (MCs) as shown in Fig. \ref{figure 1}. MCs such as laptops, desktops, mobile phones, and other wireless devices connect to the internet via MRs, which transmit traffic to and from MGs. MGs are in turn connected to the internet infrastructure.

Although WMN has certain desirable characteristics, there are a number of problems preventing their large-scale deployment. One of the critical issues receiving significant attention in the literature is the mesh router nodes placement problem, it is also known as NP-hard \cite{taleb6}, \cite{seetha7}, \cite{amaldi8}. The bad position of mesh nodes (MR and/or MG) has a significant impact on WMN performance \cite{qiu9}. Consequently, many interferences and congestion occur, leading to substantial packet loss, low throughput, and high delays. 

Several papers have presented meta-heuristic algorithms as successful solutions for solving the nodes placement problem in WMNs. Most of them considered a stationary topology, while others have focused on investigating the dynamic placement of mesh nodes  \cite{lin10}, \cite{lin11}, \cite{lin12}, \cite{Binh13}, \cite{sayad14}, \cite{oda15}. 

To address the stationary variant of the WMNs nodes place- ment problem, Xhafa et al. proposed three algorithms: Simulated Annealing (SA) \cite{xhafa16}, Hill      Climbing (HC) \cite{xhafa17}, and Tabu Search (TS) \cite{xhafa18}, and assessed their performance in terms of user coverage and network connectivity. 

Sayad et al. introduced a Chemical Reaction Optimization (CRO) algorithm \cite{sayad19}, inspired by the interactions between molecules that aim to achieve a low, stable energy state during chemical reactions. Sayad et al. also proposed a bio-inspired algorithm called the Firefly Optimization Algorithm (FA) \cite{sayad20} and compared it with the existing Genetic Algorithm. Both algorithms were tested using generated instances with varying numbers of mesh clients and routers.

Evolutionary algorithms, such as the Genetic Algorithm (GA), are also widely used for optimization in this field \cite{Xhafa21}, \cite{Tang22}, \cite{DeMarco23}, \cite{Bello24}. Xhafa et al. addressed the issue of mesh router node placement by treating it as a facility location problem and solving it using a Genetic Algorithm (GA)  \cite{Xhafa21}. Another study introduced an enhanced GA that integrates the Minimum Spanning Tree (MST) to improve cost and coverage outcomes \cite{Tang22}. In \cite{DeMarco23}, an advanced version of GA, named MOGAMESH, was developed to optimize WMN topology by maximizing user coverage and minimizing node degree. Additionally, two variations of GA were explored in \cite{Bello24}: the Non-dominated Sorting Genetic Algorithm-II (NSGA-II) and the Multi-Objective Genetic Algorithm (MOGA), which consider cost, coverage, and reliability as key performance metrics. These studies represent some of the most effective applications of multi-objective algorithms to achieve simultaneous optimization of multiple objectives in this domain \cite{Farahani25}, \cite{Zavala26}, \cite{Alothaimeen27}.

Also, recently, Mekhmoukh et al. used the known Coyote Optimization algorithm \cite{Mekhmoukh28} to solve the mesh router node placement problem, outperforming FA, PSO, BA, and other algorithms in terms of user coverage and connectivity.

Several methods have been proposed to address the dynamic variant of mesh nodes placement, as discussed in \cite{lin10}, \cite{lin11}, \cite{lin12}, \cite{sayad14}. In \cite{lin10}, an enhanced PSO algorithm incorporating a restriction coefficient into its framework was introduced to tackle this challenge. Similarly, Lin et al. \cite{lin11} presented an improved bat-inspired algorithm (BA) by inte grating a dynamic search scheme into the original BA model. This enhancement was validated through experiments on 10 instances, considering parameters such as coverage and connectivity. Authors in \cite{lin12} concentrated on the social-aware dynamic  placement of router nodes in WMNs. They introduced an enhanced PSO variant termed social-based-PSO, which incorporates a social-supporting vector. 

In this work, we propose a new algorithm for mesh router nodes placement based on Entropy and Genetic Algorithm. We assess the performance of MEGA through numerous simulations with different settings, considering both coverage and connectivity metrics. Our approach is inspired by GA, which is known for its robust optimization capabilities. GA imitates the process of natural selection, where the fittest individuals are selected for reproduction in order to produce the offspring of the next generation. The primary advantages of GA include its ability to efficiently search large and complex spaces, its flexibility in handling various types of objective functions, and its robustness against getting trapped in local optima.

In our proposed algorithm, the fitness function is calculated using Shannon's entropy, aiming to get the maximum entropy value \cite{Shannon29}. According to this theory, the entropy is maximized when the probability distribution of the nodes is uniform. This method of calculating the fitness function through entropy ensures a uniformly distributed mesh routers nodes placement considering mesh clients positions.

The rest of the paper is organized as follows. Section \ref{sec:problem formulation} details the formulation of the mesh router nodes placement problem. In Section \ref{sec:MEGA}, we introduce a novel Entropy and GA inspired algorithm (MEGA), designed to address the mesh router nodes placement problem.    Section \ref{sec:results} contains simulation results and comparison with other approximate optimization algorithms. Finally, conclusion is given in Section \ref{sec:conclusion}.

\section{Mesh nodes placement problem formulation}
\label{sec:problem formulation}
In this section, we propose the system model and formulate the problem regarding the placement of mesh router nodes. For better readability, the notations used in this paper are presented in Table \ref{tabel 1}.

\subsection{System model}
WMN can be mathematically represented as an undirected graph $G = (V, E)$ where $V$ represents the set of network nodes and $E$ denotes the links connecting these nodes. The network $G$ comprises several disjoint subnetworks. In this work, the WMN includes two types of nodes: mesh clients and mesh routers. Thus, $V = MR \cup MC$ where:

\begin{itemize}
    \item $MR$ is the set of $m$ mesh routers, denoted as $MR = \{mr_1, mr_2, \ldots, mr_m\}$. Each router is equipped with a radio interface having the same coverage radius, denoted as $CR_1 = CR_2 = \ldots = CR_m$. Two mesh routers, $mr_i$ and $mr_j$, can connect only if the distance between them, $d(mr_i, mr_j)$, is less than or equal to twice the coverage radius $CR$, i.e., $d(mr_i, mr_j) \leq 2CR$.
    
    \item $MC$ is the set of $n$ mesh clients, represented as $MC = \{mc_1, mc_2, \ldots, mc_n\}$. Here, mesh clients are randomly distributed within a two-dimensional rectangular area with dimensions $W \times H$. A mesh client $mc_i$ is considered covered by a mesh router $mr_j$ if it falls within the router’s coverage radius, i.e., $d(mc_i, mr_j) \leq CR$. Each client can be associated with only one router, typically the nearest one, although it may be within the coverage radius of multiple routers.
\end{itemize}

\begin{table}[h]
\centering
\caption{\textbf {The main notations used in this paper.}}
\begin{tabular}{|c|l|}
\hline
\textbf{Parameter} & \textbf{Description}                           \\ \hline
$G = (V, E)$       & Undirected graph                               \\ \hline
$V$                & Set of mesh nodes                              \\ \hline
$E$                & Set of links between mesh nodes                \\ \hline
$MR$               & Set of mesh routers                            \\ \hline
$MC$               & Set of mesh clients                            \\ \hline
$m$                & Number of mesh routers                         \\ \hline
$n$                & Number of mesh clients                         \\ \hline
$mr_i$             & The $i$-th mesh router                         \\ \hline
$mc_i$             & The $i$-th mesh client                         \\ \hline
$CR_i$             & Coverage radius of the $i$-th mesh router      \\ \hline
$G_i = (V_i, E_i)$ & The $i$-th sub-network                         \\ \hline
$|G_i|$            & Size of the $i$-th sub-network                 \\ \hline
$ G_n $            & Number of sub-networks                         \\ \hline
$\phi(G)$          & Network connectivity                           \\ \hline
$\psi(G)$          & User coverage                                  \\ \hline
$W$                & Width of the dimension                         \\ \hline
$H$                & Height of the dimension                        \\ \hline
$n_j$              & Number of covered clients in j-th router       \\ \hline
$P_i$              & Probability of coverage                        \\ \hline
$P_j$              & Probability of connectivity                    \\ \hline
$H_{\text{cov}}$   & Coverage entropy                               \\ \hline
$H_{\text{con}}$   & Connectivity entropy                           \\ \hline
\end{tabular}
\label{tabel 1}
\end{table}

\subsection{Problem formulation}
Depending on the nature of the environments studied (static or dynamic) and the type of deployment spaces (discrete or continuous), various variants of the WMN router nodes placement problem can be identified. In this paper, we focus on the static continuous placement of mesh routers. The primary objective is to determine the optimal positioning of $m$ mesh routers within a two-dimensional area with dimensions $W \times H$, taking into account the positions of $n$ mesh clients \cite{oda30}, \cite{benyamina31}, \cite{Seetha32}. 

The problem in this article aims to optimize two main objectives:

\begin{itemize}
    \item \textbf{User coverage:} This refers to the count of users covered by at least one mesh router and can be found according to the formula \cite{Mekhmoukh28}:
    \begin{equation}
    \Psi (G) = \sum_{i=1}^{n} \left( \max_{j \in \{1, \ldots, m\}} \sigma_{ij} \right)
    \label{eq:1}
    \end{equation}

    where $\sigma_{ij}$ represents the coverage variable, defined as:
    
    \begin{equation}
    \sigma_{ij} = 
    \begin{cases} 
    1 & \text{if mesh client $c_i$ is covered by mesh router $r_j$,} \\
    0 & \text{otherwise.}
    \end{cases}
    \label{eq:2}
    \end{equation}

    \item \textbf{Network connectivity:} This is defined as the largest sub-network among $k$ formed sub-networks considering the number of mesh nodes (both routers and clients). It can be found as \cite{Mekhmoukh28}:
    \begin{equation}
    \Phi (G) = max_{i \in \{1, \ldots, k\}} |G_i|   
    \label{eq:3}
    \end{equation}
    where $|G_i|$, for $i \in \{1, k\}$, denotes the size of the $i^{th}$ sub-network, and $G = G_1 \cup G_2 \cup \ldots \cup G_k$.
\end{itemize}

\section{Maximum entropy genetic algorithm}
\label{sec:MEGA}
In this section, we present a detailed explanation of MEGA for mesh router nodes placement. Our methodology consists of two parts: GA and Entropy Fitness Estimation.
\subsection{Genetic Algorithm}
\begin{figure}
    \centering
    \includegraphics[width=1 \linewidth]{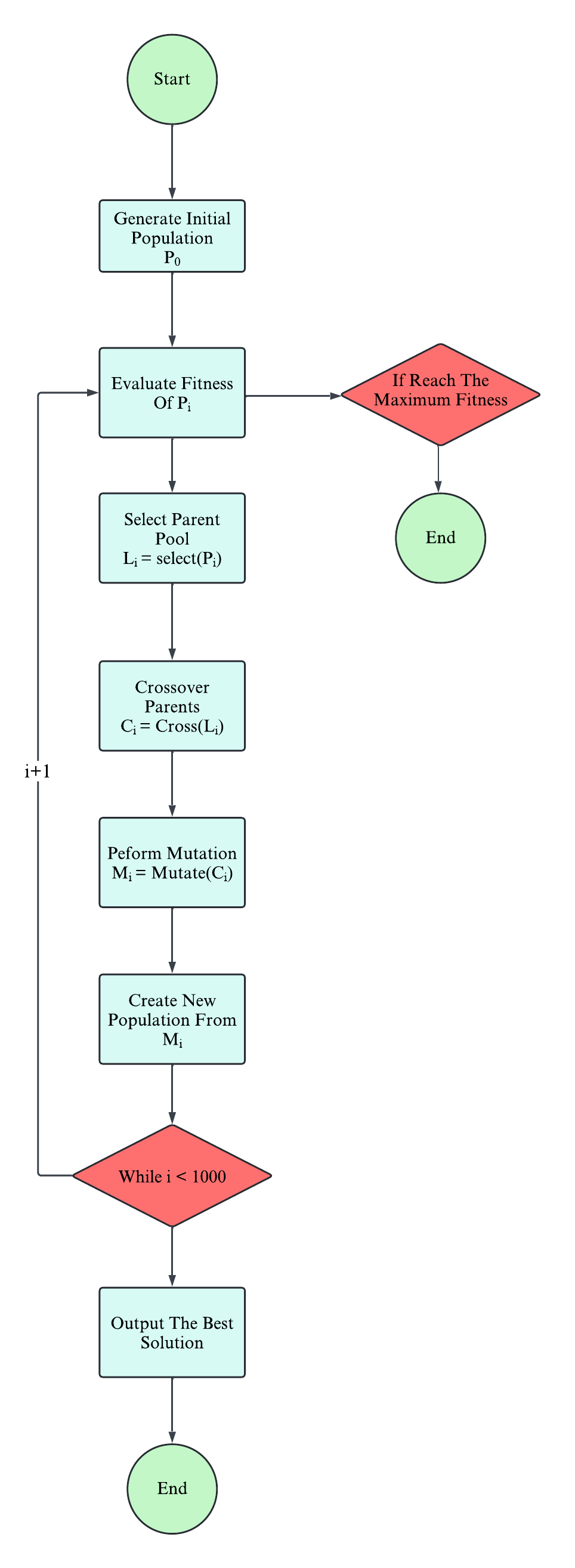}
    \caption{Flowchart of the MEGA.}
    \label{figure 2}
\end{figure}

GAs are adaptive heuristic search algorithms rooted in the principles of natural selection and genetics.    This simulation  imitate the process of evolution, where individuals represents potential solutions compete for resources and opportunities to reproduce \cite{mitchell33,goldberg34}. Through selection, crossover, and mutation, GA iteratively refines the population, favoring individuals with higher fitness.    This emulation of "survival of the fittest" leads to the generation of high-quality solutions for optimization and search problems\cite{holland35,eiben36}. The flowchart illustrating the MEGA algorithm is presented in Fig. \ref{figure 2}.

\begin{itemize}
    \item \textit{Initialization:} At the beginning, clients are uniformly randomly distributed within two dimensional area. Then, to initiate the optimization process, a random set of candidate solutions, specifically mesh routers, is generated. This step involves initializing a population of a specific size, where each individual is represented by a chromosome. The chromosome in our method indicates the positions of mesh routers, with its length matching to the number of routers within the clients' distributed area. After that, the fitness functions of each router within the population are calculated, which will be used for parents selection. A detailed explanation of how fitness is calculated is presented in the next subsection.
    
    \item \textit{Selection:} After evaluating the fitness of the population, we proceed with the selection operator to identify the top-performing individuals for reproduction. This implies selecting the highest 20 percent of the population based on their fitness values. By sorting the fitness values and selecting the corresponding individuals, we define this subset as parents for the next generation.
    
    \item \textit{Crossover operators:} The crossover operator plays a crucial role in GAs, facilitating the transmission of advantageous genetic traits to future generations and driving evolutionary progress. In our implementation, the crossover point is randomly selected within the length of the chromosome, represented as arrays of coordinates. This diversifies solutions, enhancing the genetic algorithm's effectiveness in optimizing mesh router nodes placement.
    
    \item \textit{Mutation operators:} Mutation operators in GAs typically lead to minor local changes of individuals' chromosome, contrasting with crossover operators. The mutation process introduces variability by randomly altering individual genes within the chromosome. Each gene has a probability of being mutated, determined by the adaptive mutation rate. The likelihood of mutation decreases as fitness approaches its maximum value, leading to optimal solutions. 
    
    \item \textit{Optimal Result Output:} Following mutation, the algorithm reevaluates the fitness of the mutated individuals. It iterates through the selection, crossover, and mutation steps until the specified number of iterations is reached or until it achieves the maximum fitness value, outputting the optimal result.
\end{itemize}

\subsection{Entropy fitness estimation}
In our proposed method the fitness function is calculated based on Shannon entropy \cite{Shannon29}.  Entropy is a fundamental concept in information theory that quantifies uncertainty and probability, providing insight into the information content within a system\cite{csiszar37,cover38}. In our algorithm, information includes both the uniform distribution of covered clients and the interconnectivity among mesh routers. These aspects are used for estimating the fitness function by defining connectivity and coverage entropy based on the network topology. The coverage entropy evaluates how covered clients are dispersed by the mesh routers, considering uncertainty in client coverage within the network. Similarly, the connectivity entropy quantifies uncertainty in the interconnections among mesh routers and clients. 

The coverage entropy (\( H_{\text{cov}} \)) can be calculated according to the following formula:
\begin{equation}
H_{\text{cov}} = -\frac{{\sum_{i}^{m} P_i \ln(P_i)}}{\ln(m)},  
\label{eq:4}
\end{equation} where \( m \) indicates the total number of mesh routers, \( P_i \) denotes coverage probability. The entropy calculation includes iterating over each mesh router's position and verifying the distance to each clients. If a client falls within the specified covering radius of the router, it increase a coverage count \( n_j \).This count is then divided  by the total number of clients \( n \) to determine the coverage probability  \( P_i \). Dividing by \( ln(m) \) in the calculation normalizes the entropy value, ensuring that the fitness function of the optimal solution approaches 1. Below in  Fig. \ref{figure 3} we provide a detailed explanation of \( H_{\text{cov}} \). In Fig. \ref{figure 3} each router covers the same number of clients and the  \( H_{\text{cov}} \) reaches its maximum. In this scenario, the probability distribution is obviously equal.

 \begin{figure}[h!]
    \centering
    \includegraphics[width=0.8\linewidth]{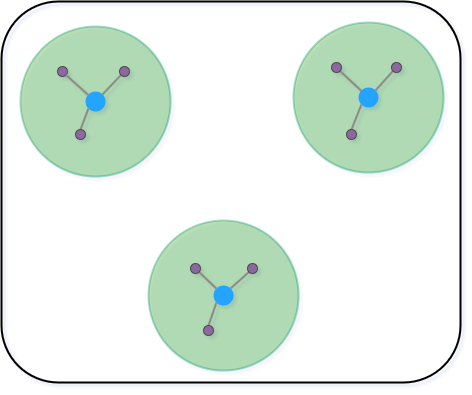} 
    \caption{\textbf{ The scenario illustrates an equal coverage probability distribution, where blue nodes represent mesh routers (3 units) and purple nodes represent mesh clients (9 units). Each  \( P_i \)  is equal to $\frac{1}{3}$, resulting \( H_{\text{cov}} \) = 1.}}
    \label{figure 3}
\end{figure}

\begin{figure}[h!]
    \centering
    \includegraphics[width=0.8\linewidth]{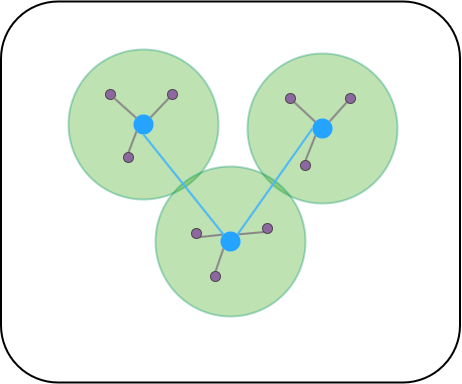} 
    \caption{\textbf{The scenario illustrates case when \( H_{\text{con}} \) = 0, resulting in best connectivity among nodes. Here, blue nodes represent mesh routers, purple nodes represent mesh clients, and blue lines represent the connectivity between mesh routers.}}
    \label{figure 4}
\end{figure}

The connectivity entropy (\( H_{\text{con}} \)) can be calculated according to the following formula : 

\begin{equation}
    \begin{cases}
        H_{\text{con}} = -\frac{{\sum_{j}^{G_n} P_j \ln(P_j)}}{\ln(G_n)} ; G_n > 1 \\ 
        H_{\text{con}} = 0 ; G_n = 1,
    \end{cases}
\label{eq:5}
\end{equation} where \( G_n \) indicates the number of sub-networks, \( P_j \) represents connection probability. We establish connectivity between routers when their Euclidean distance falls within twice the \( CR \), defining these connected mesh routers and clients within the  \( CR \) as components as \( |G_i| \).  We then calculate \( P_j \) by dividing   \( |G_i|\) by the total count of  clients and  routers. After that, we calculate \( G_n \) ,which represents each cluster of interconnected nodes. It acts as a normalization factor to reduce \( H_{\text{con}} \) toward 0. Notably, when the count of  \( G_n \) equals 1, indicating the connection of all components,  \( H_{\text{con}} \) becomes 0 (Fig. \ref{figure 4}).

The final fitness function is derived from \( H_{\text{cov}} \) - \( H_{\text{con}} \). As \( H_{\text{cov}} \) approaches 1 and \( H_{\text{con}} \) tends towards 0, the fitness function converges 1, reflecting an optimal network configuration.

\section{Results and discussion}
\label{sec:results}
In this section, we evaluate the performance of the proposed MEGA algorithm for addressing the mesh router nodes placement problem in WMNs. The MEGA algorithm is compared with four top-performing methods such as FA \cite{sayad20}, GA \cite{oda15}, PSO \cite{lin10}, and COA as discussed by   Mekhmoukh et al. \cite{Mekhmoukh28}. 

We assess these algorithms based on three key performance metrics: user coverage, network connectivity, and the value of the objective fitness function. MEGA is implemented using Python environment. All tests are conducted using a Core i7 5.2 GHz CPU machine. Simulations are carried out in a rectangular area measuring 2000 m x 2000 m. The number of mesh routers tested varies between 5 and 40, aimed at  operating 50 to 300 mesh clients, which are randomly positioned within the test area. Each set of tests includes 1000 iterations, and the results are the average outcomes from 50 trials. 
The simulation parameters are given in  Table \ref{table 2}. Our research investigates the impact of different variables, such as the number of mesh clients and routers, as well as the coverage radius. 

\begin{table}[ht]
\centering
\caption{\textbf {Parameters values considered in our simulations.}}
\begin{tabular}{|c|l|l|}
\hline
\textbf{Parameter} & \textbf{Value}           & \textbf{ Default value}      \\ \hline
$n$                & [50, 300]                & 100                          \\ \hline
$m$                & [5, 40]                  & 20                           \\ \hline
$CR$               & [50, 400] m               & 200 m                         \\ \hline
$W$                & 2000 m                      & 2000 m                         \\ \hline
$H$                & 2000 m                     & 2000 m                         \\ \hline
$Population$ $size$  & 50                       & 50                           \\ \hline
$Number$ $of$ $runs$    & 50                       & 50                           \\ \hline
$Number$ $of$ $iterations$  & 1000                 & 1000                          \\ \hline
\end{tabular}
\label{table 2}
\end{table}

\begin{figure}[htbp!]
    \centering
    \includegraphics[width=1\linewidth]{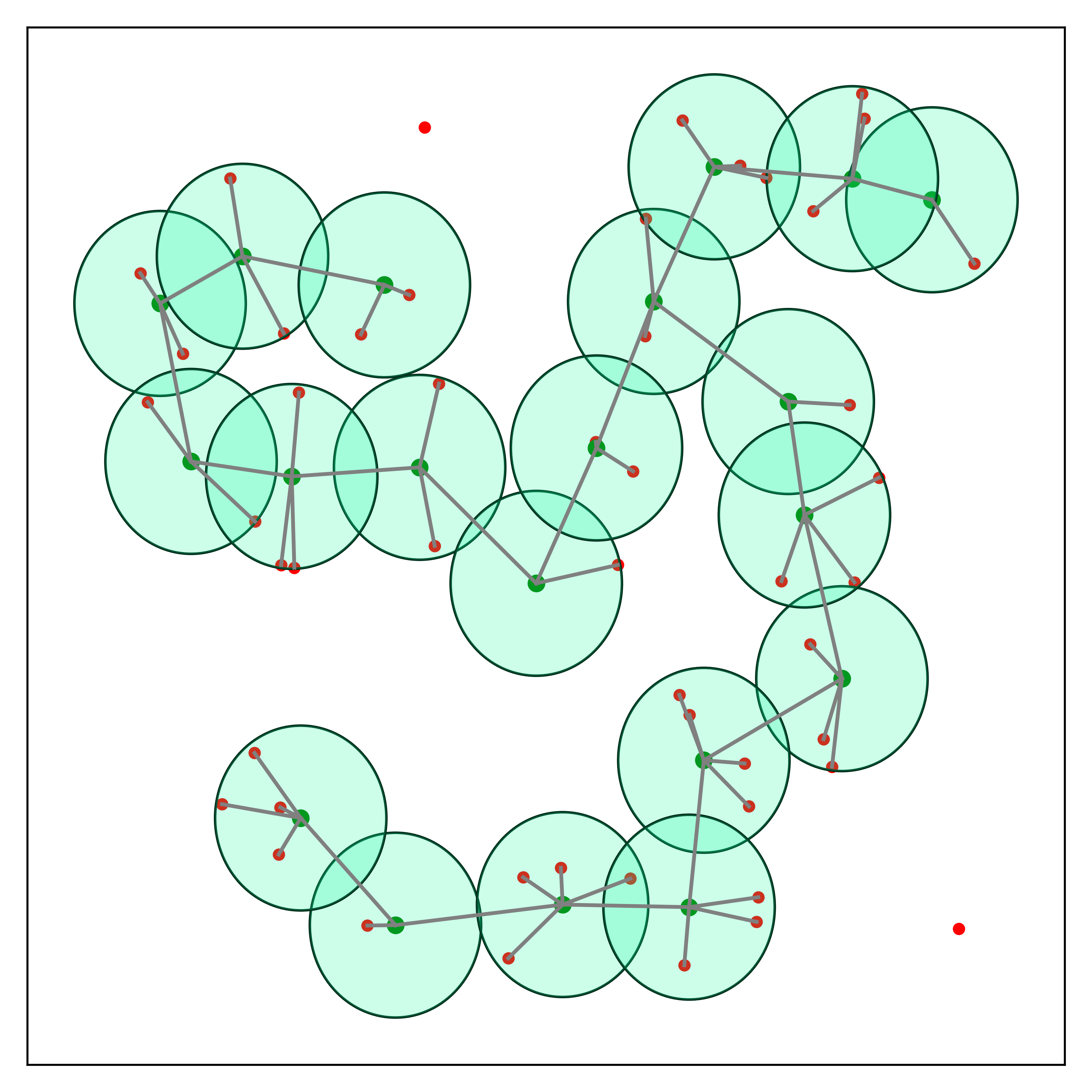} 
    \caption{\textbf{The optimal placement of mesh routers obtained using MEGA. Green nodes denote mesh routers, red nodes indicate mesh clients, and lines between routers show connectivity.}}
    \label{figure 5}
\end{figure}

Fig. \ref{figure 5} shows example of a planned WMNs using MEGA, with network designed for scenario representing 20 mesh routers and 50 clients uniformly distributed over 4$ \, \text{km}^2$ area.

\subsection{IMPACT OF VARYING THE NUMBER OF MESH CLIENTS}
\begin{figure}[htbp!]
    \centering
    \includegraphics[width=1\linewidth]{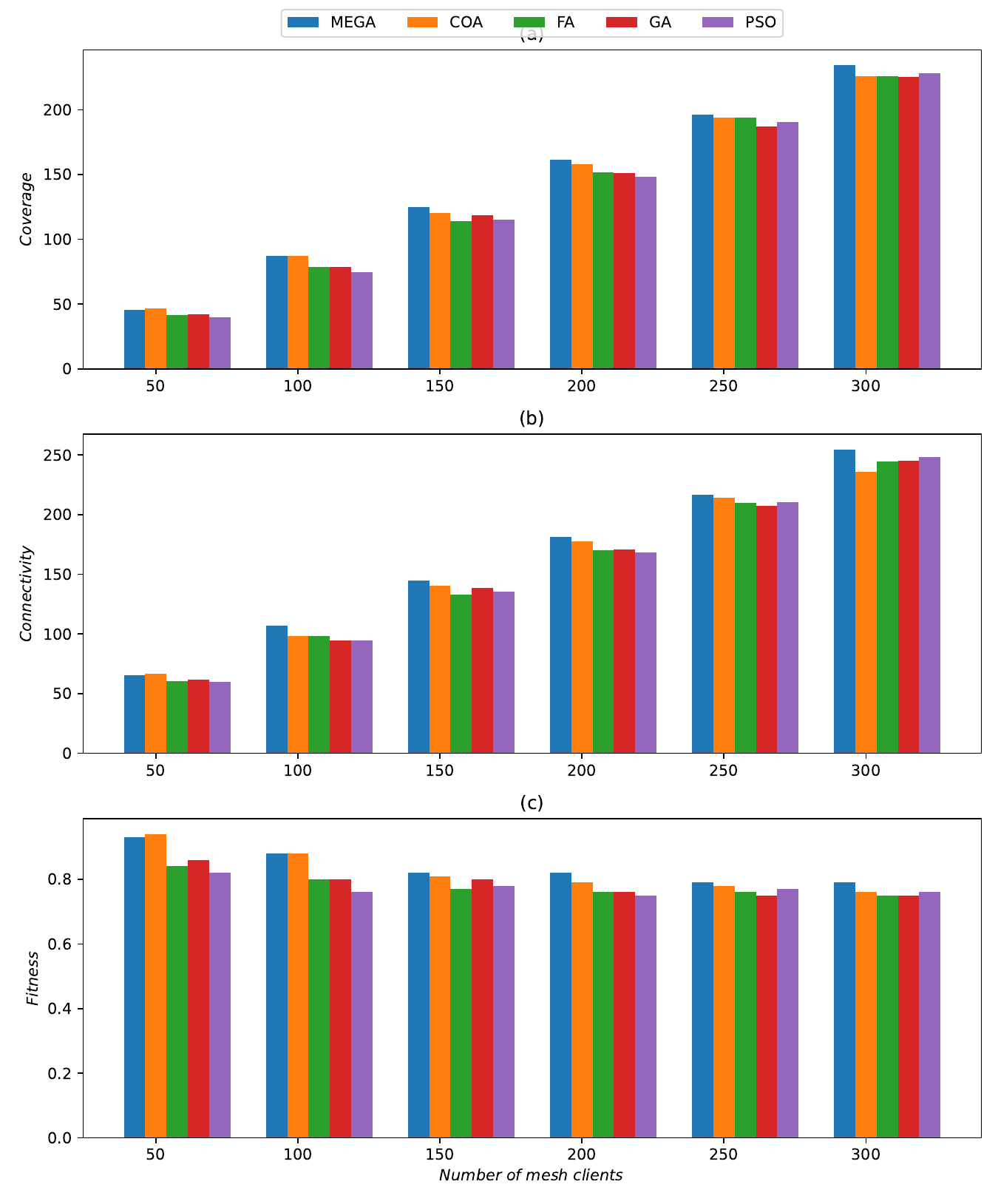} 
    \caption{\textbf{Impact of varying number of mesh clients on: (a) Coverage (b) Connectivity (c) Fitness.}}
    \label{figure 6}
\end{figure}

In this scenario, we varied the number of mesh clients from 50 to 300 while the number of mesh routers was constant.  Table \ref{table 3} details the influence of increasing the number of mesh clients on user coverage, network connectivity, and the fitness function. In Fig. \ref{figure 6}, its graphical representation is given. In Table \ref{table 3} the data of COA, FA, GA and PSO algorithms is taken from \cite{Mekhmoukh28}. Fig. \ref{figure 6} (a) illustrates the variation in user coverage as the number of mesh clients increases. We observed a consistent increase in user coverage with increasing number of clients. Also, our method demonstrates better performance in client coverage compared to other alternatives: 1.5\% more than COA, 7\% more than FA, 6.7\% more than GA and 9.2\% more than PSO.

\begin{table}[h!]
\centering
\caption{\textbf{Coverage, connectivity, fitness under various numbers of mesh clients.}}
\begin{tabular}{|l|c c c c c c|}
\cline{1-7}
\textbf{n} & \textbf{50} & \textbf{100} & \textbf{150} & \textbf{200} & \textbf{250} & \textbf{300} \\
\cline{1-7}
\multicolumn{7}{|c|}{\textbf{Coverage}} \\
\cline{1-7}
MEGA  & 45.7 & 87 & \textbf{125} & \textbf{161.4} & \textbf{196.5} & \textbf{234.6} \\ 
COA  & \textbf{46.74} & \textbf{87.12} & 120.32 & 157.88 & 194 & 226 \\ 
FA   & 41.44 & 78.74 & 113.82 & 151.63 & 194 & 222.16 \\
GA   & 42.26 & 78.6  & 118.86 & 151.1  & 187.13 & 225.33 \\
PSO  & 39.63 & 74.56 & 115.33 & 148.53 & 190.66 & 228.03 \\
\cline{1-7}
\multicolumn{7}{|c|}{\textbf{Connectivity}} \\
\cline{1-7}
MEGA  & 65.7 & 107 & \textbf{145} & \textbf{181.4} & \textbf{216.5} & \textbf{254.6} \\ 
COA  & \textbf{66.5}  & \textbf{107.12} & 140.32 & 177.5  & 214 & 248.2 \\
FA   & 60.38 & 98.24  & 132.8  & 170.1  & 210.16 & 244.84 \\
GA   & 62    & 98.6   & 138.86 & 171.1  & 207.13 & 245.33 \\
PSO  & 59.63 & 94.4   & 135.33 & 168.53 & 210.66 & 248.03 \\
\cline{1-7}
\multicolumn{7}{|c|}{\textbf{Fitness}} \\
\cline{1-7}
MEGA  & 0.93 & \textbf{0.88} & \textbf{0.82} & \textbf{0.82} & \textbf{0.79} & \textbf{0.79} \\ 
COA  & \textbf{0.94} & 0.88 & 0.81 & 0.79 & 0.78 & 0.76 \\
FA   & 0.84 & 0.80 & 0.77 & 0.76 & 0.75 & 0.75 \\
GA   & 0.86 & 0.80 & 0.80 & 0.76 & 0.76 & 0.75 \\
PSO  & 0.82 & 0.76 & 0.78 & 0.75 & 0.77 & 0.76 \\
\cline{1-7}
\end{tabular}
\label{table 3}
\end{table}

Fig. \ref{figure 6} (b) illustrates that network connectivity improves as the number of mesh clients increases. It is demonstrated that our method significantly increased network connectivity. More specifically, connectivity is improved on average by 3.71\%, 6.7\%, 6.34\% and 7.3\%  compared to COA, FA, GA and PSO, respectively. Results shown in Fig. \ref{figure 6} (c) illustrates a decline in fitness values as the number of mesh clients increases, necessitating more routers to keep coverage. With a fixed number of mesh routers, newly added clients might not be covered, resulting in reduced coverage and connectivity, which impacts the fitness value. The obtained results revealed that MEGA performs better than COA, FA, GA and PSO.

\subsection{IMPACT OF VARYING THE NUMBER OF MESH ROUTERS}
\begin{figure}[htbp!]
    \centering
    \includegraphics[width=1\linewidth]{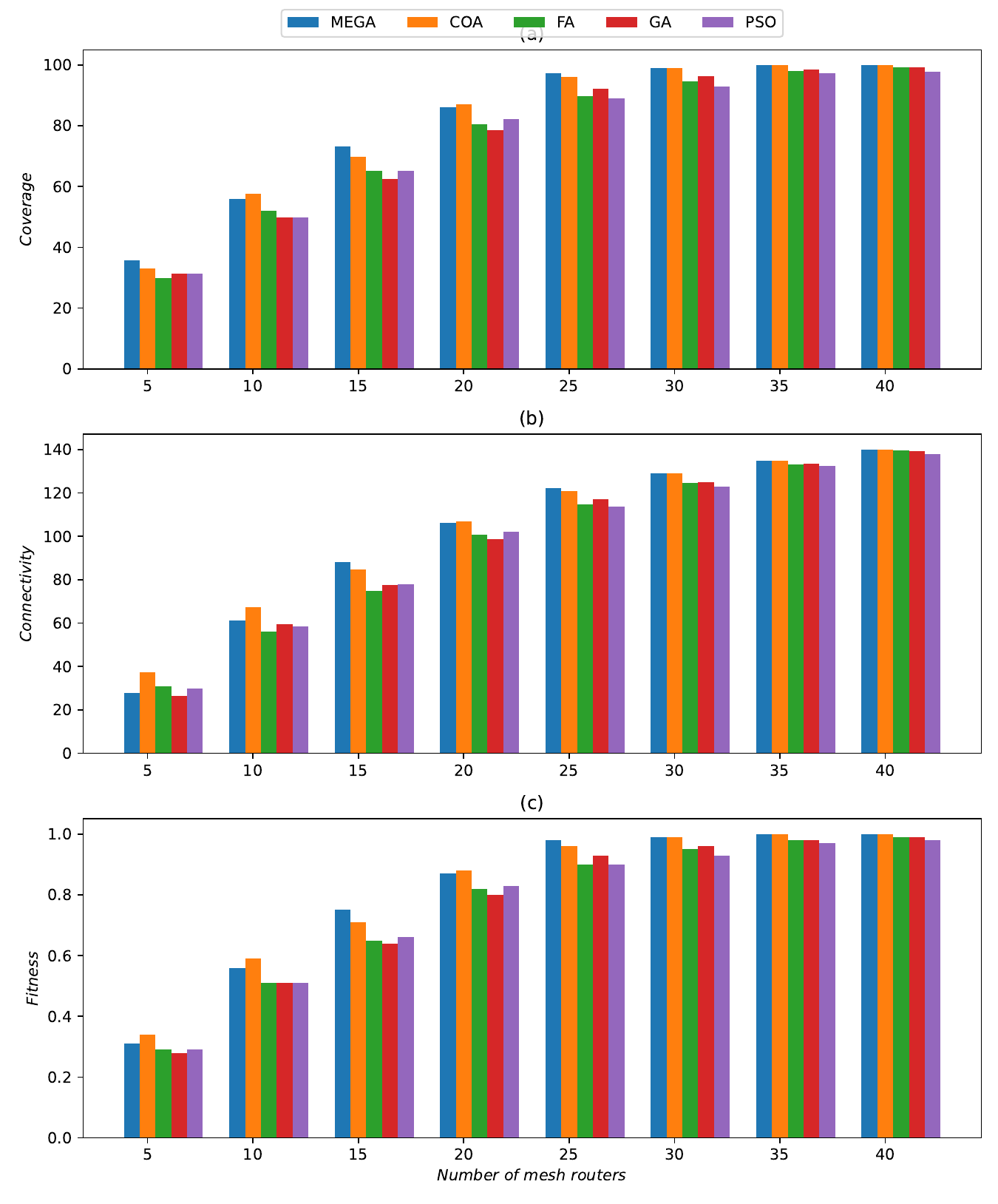} 
    \caption{\textbf{Impact of varying number of mesh routers on: (a) Coverage (b) Connectivity (c) Fitness.}}
    \label{figure 7}
\end{figure}

The influence of varying the number of mesh routers (from 5 to 40) on coverage, connectivity, and overall fitness value are given in  Table \ref{table 4} and Fig. \ref{figure 7}. In Table \ref{table 4} the data of COA, FA, GA and PSO algorithms is taken from \cite{Mekhmoukh28}. As depicted in Fig. \ref{figure 7} (a), the coverage of users improves with an increase of mesh routers. More specifically, the coverage is improved on average by 1.7\%, 9.9\% , 10.2\% and 9.8\% in comparison to the COA, FA, GA, and PSO algorithms, respectively. Fig. \ref{figure 7} (b) illustrates that network connectivity also rises with an increase of mesh routers. This increase results from the reduction in the number of isolated subnetworks as additional routers help to eliminate gaps, forming larger sub-networks. Finally, this leads to the formation of a single, extensive subnetwork encompassing all mesh nodes. The MEGA algorithm achieves the largest subnetwork compared to others, with average connectivity improvements of  5.4\%, 6.2\% and 4.5\% over FA, GA and PSO, excluding COA which performed 5.2\% better.

\begin{table}[h!]
\centering
\setlength{\tabcolsep}{4pt}
\caption{\textbf{Coverage, connectivity, fitness under various number of mesh routers.}}
\begin{tabular}{|l|c c c c c c c c|}
\cline{1-9}
\textbf{m} & \textbf{5} & \textbf{10} & \textbf{15} & \textbf{20} & \textbf{25} & \textbf{30} & \textbf{35} & \textbf{40} \\
\cline{1-9}
\multicolumn{9}{|c|}{\textbf{Coverage}} \\
\cline{1-9}
MEGA  & \textbf{35.6} & 55.9 & \textbf{73.3} & 86.2 & \textbf{97.2} & \textbf{99.1} & \textbf{100} & \textbf{100}\\ 
COA  & 32.96 & \textbf{57.7} & 69.9 & \textbf{87} & 96 & 99 & 100 & 100\\ 
FA   & 29.86 & 52 & 65.2 & 80.6 & 89.8 & 94.6 & 98.06 & 99.36 \\
GA   & 31.23 & 49.93  & 62.43 & 78.6  & 92.23 & 96.26 & 98.53 & 99.36\\
PSO  & 31.33 & 49.86 & 65.2 & 82.33 & 89.06 & 92.86 & 97.33 & 97.86\\
\cline{1-9}
\multicolumn{9}{|c|}{\textbf{Connectivity}} \\
\cline{1-9}
MEGA  & 27.7 &	61.2 &	\textbf{88.3} &	106.2 &	\textbf{122.2} &	\textbf{129.1} &	\textbf{135} &	\textbf{140} \\ 
COA  & \textbf{37.42} &	\textbf{67.43} &	84.9 &	\textbf{107} &	121	& 129 & 	135 &	140\\
FA   & 30.86&	56&	75&	100.6&	114.8&	124.6&	133	&139.76 \\
GA   & 26.46&	59.43&	77.43	&98.6&	117.23&	124.8&	133.53&	139.36 \\
PSO  &29.86  &	58.58  &	77.96  &	102.2  &	113.8  &	122.86  &	132.33  &	137.86 \\
\cline{1-9}
\multicolumn{9}{|c|}{\textbf{Fitness}} \\
\cline{1-9}
MEGA  & 0.31&	0.56&	\textbf{0.75}&	\textbf{0.88}&	\textbf{0.98}& \textbf{0.99}&	\textbf{1}	&\textbf{1} \\ 
COA  & \textbf{0.34}&	\textbf{0.59}&	0.71&	0.88	&0.96&0.99	&1	&1 \\
FA   & 0.29&	0.51&	0.65&	0.82&	0.9&0.95&	0.98&	0.99 \\
GA   & 0.28&	0.51&	0.64&	0.8&	0.93&	0.96&	0.98&	0.99 \\
PSO  & 0.29&0.51&	0.66&	0.83&	0.9&0.93&	0.97&	0.98 \\
\cline{1-9}
\end{tabular}
\label{table 4}
\end{table}

Fig. \ref{figure 7} (c) demonstrates that the fitness value correlates positively with the number of mesh routers. As the number of routers increases, the fitness value improves across all algorithms. The proposed MEGA consistently surpasses COA, FA, GA and PSO in performance when the number of mesh routers exceeds 10. This trend suggests that increasing router density not only enhances coverage and connectivity but also significantly enhances overall network performance.

\subsection{IMPACT OF VARYING THE ROUTER COVERAGE RADIUS}
\begin{figure}[h]
    \centering
    \includegraphics[width=1\linewidth]{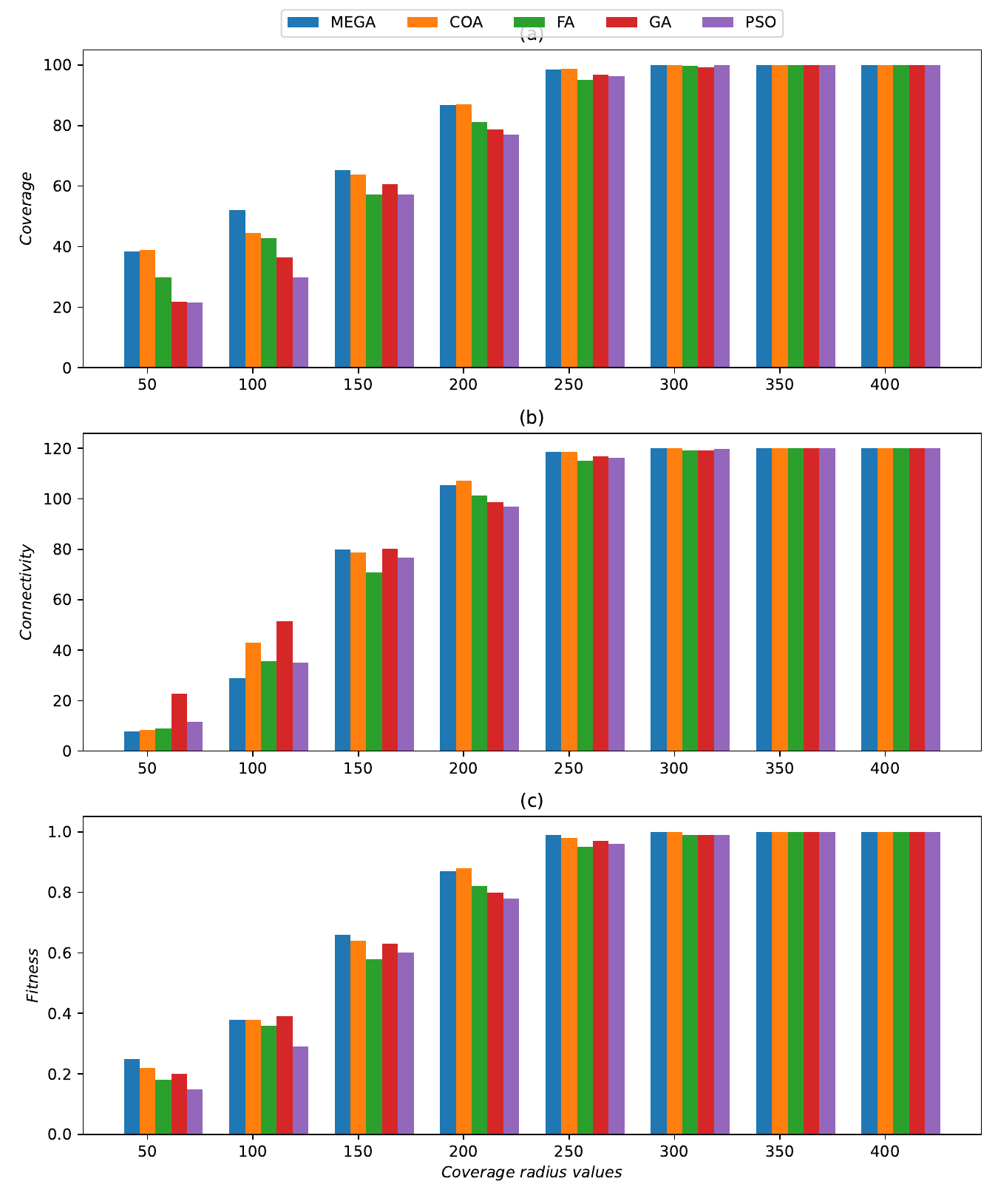} 
    \caption{\textbf{Impact of varying coverage radius values on: (a) Coverage (b) Connectivity (c) Fitness.}}
    \label{figure 8}
\end{figure}

 Table \ref{table 5} and Fig. \ref{figure 8} illustrate the influence of varying the coverage radius of mesh routers from 50 to 400 m on coverage, connectivity, and fitness values. In Table \ref{table 5} the data of COA, FA, GA and PSO algorithms is taken from \cite{Mekhmoukh28}. Fig. \ref{figure 8} (a) shows the impact of expanding the coverage radius on network coverage. The results indicate that as each mesh router's coverage radius is extended, there is a corresponding increase in the coverage metric. Specifically, when the coverage radius exceeds 300 m, most routers can cover almost all mesh clients. The MEGA exceeds the performance of other algorithms across all scenarios, improving client coverage on average to 2.8\%, 12.5\%, 23.4\% and 30\% compared to COA, FA, GA and PSO, respectively.

\begin{table}[h!]
\centering
\setlength{\tabcolsep}{4pt}
\caption{\textbf{Coverage, connectivity, fitness under various coverage radius values.}}
\begin{tabular}{|l|c c c c c c c c|}
\cline{1-9}
\textbf{CR} & \textbf{50} & \textbf{100} & \textbf{150} & \textbf{200} & \textbf{250} & \textbf{300} & \textbf{350} & \textbf{400} \\
\cline{1-9}
\multicolumn{9}{|c|}{\textbf{Coverage}} \\
\cline{1-9}
MEGA  & 38.3&	\textbf{52.1}&	\textbf{65.2}&	86.8&	98.6&	\textbf{100}	&\textbf{100}&	\textbf{100}\\ 
COA  & \textbf{38.88}&	44.6&	63.78& \textbf{87.12}&	\textbf{98.72}&100	&100&100\\ 
FA   & 29.9&	42.76&	57.2&	81.2&	95.13&	99.66&	100	&100 \\
GA   & 21.73&	36.36&	60.53	&78.6&	96.76&	99.16&100&	100\\
PSO  & 21.63&29.9	&57.1&	76.93&	96.26&	99.86&	100	&100\\
\cline{1-9}
\multicolumn{9}{|c|}{\textbf{Connectivity}} \\
\cline{1-9}
MEGA  & 7.9&29	&\textbf{80}	&105.5&	118.6&	\textbf{120}	&\textbf{120}&	\textbf{120} \\ 
COA  & \textbf{8.26}&	\textbf{42.88}&	78.72&	\textbf{107.12}& \textbf{118.72}&	120&	120	&120\\
FA   & 9	&35.73	&70.82&	101.2&	115.13	&119.2&	120&120 \\
GA   & 22.63&	51.63&	80.1&	98.6&	116.76	&119.16&	120&	120 \\
PSO  &11.63&34.96&	76.63&	96.93&	116.26&119.86&120&120 \\
\cline{1-9}
\multicolumn{9}{|c|}{\textbf{Fitness}} \\
\cline{1-9}
MEGA  & \textbf{0.25}&	0.38&	\textbf{0.66}	&0.87&\textbf{0.99}&	\textbf{1}&	\textbf{1}&	\textbf{1} \\ 
COA  & 0.22&	0.38&	0.64&	\textbf{0.88}&	0.98&	1&	1&	1 \\
FA   & 0.18&0.36&	0.58&	0.82&	0.95&	0.99&	1&	1 \\
GA   & 0.2&\textbf{0.39}&	0.63&	0.8&	0.97&	0.99&	1&	1 \\
PSO  & 0.15&0.29&	0.6	&0.78&	0.96&	0.99&	1&	1 \\
\cline{1-9}
\end{tabular}
\label{table 5}
\end{table}

Fig. \ref{figure 8} (b) investigates how network connectivity is influenced by the router's coverage radius. As the coverage radius of each router increases, the connectivity of the network also increases. This enhancement occurs because each router can cover more clients and establish connections with other routers, thus expanding the largest subnetwork.

Fig. \ref{figure 8} (c) indicates that the fitness value improves as the mesh router coverage radius increases. It is demonstrated that the MEGA surpasses other algorithms in enhancing fitness values across different coverage radius settings. More precisely, the fitness value is improved on average to 2.7\%, 11.6\%, 6.5\% and 20\% over COA, FA, GA and PSO, respectively.

\section{Conclusion}
\label{sec:conclusion}
In this work, we have proposed MEGA, a new algorithm to tackle the mesh router nodes placement problem.The performance of MEGA was thoroughly assessed by varying the quantity of mesh clients, mesh routers, and the values of the coverage radius. The results demonstrates that MEGA performs better than other optimization algorithms like COA, FA, GA and PSO in achieving better network connectivity and user coverage. Future research will apply the MEGA to address challenges of gateway placement, antenna positioning, routing, and channel allocation.


\bibliographystyle{unsrt}
\bibliography{refs.bib}

\begin{IEEEbiography}[{\includegraphics[width=1in,height=1.25 in,clip,keepaspectratio]{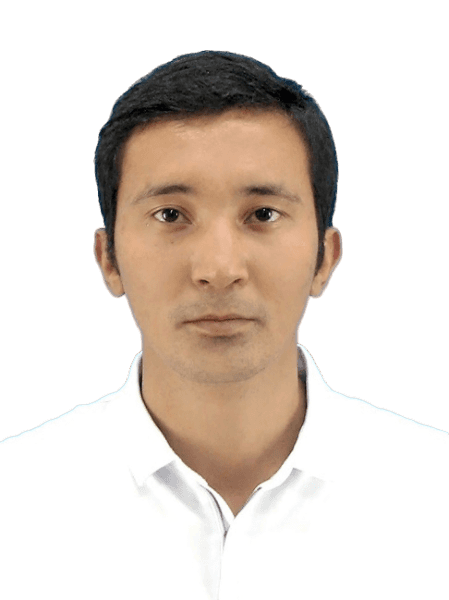}}]{N. Ussipov} received the master’s degree in engineering sciences from Taraz State University named after M. Kh. Dulaty, Taraz, Kazakhstan, in 2016.  He is currently pursuing the Ph.D degree with Al-Farabi Kazakh National University. He is a Senior Researcher and a Chief Programmer with the Department of Solid State Physics and Nonlinear Physics, Al-Farabi Kazakh National University. His current research interests include signal modulation classification, signal processing, network implementation, network information theory, and routing algorithms. 
\end{IEEEbiography}

\begin{IEEEbiography}[{\includegraphics[width=1in,height=1.25in,clip,keepaspectratio]{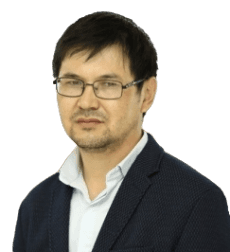}}]{S. Akhtanov} received the Ph.D. degree in physics from Al-Farabi Kazakh National University, Almaty, Kazakhstan, in 2019. He is currently a Senior Researcher with the Department of Solid State Physics and Nonlinear Physics, Al-Farabi Kazakh National University. His current research interests include signal modulation classification, signal processing, network implementation, and optimization algorithms in wireless networks. 
\end{IEEEbiography}

\begin{IEEEbiography}[{\includegraphics[width=1in,height=1.25in,clip,keepaspectratio]{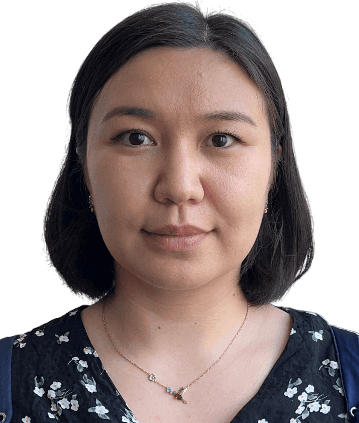}}]{D. Turlykozhayeva} received the master’s degree in physics from Tomsk Polytechnical University, Tomsk, Russia, in 2017.  She is currently pursuing the Ph.D degree with Al-Farabi Kazakh National University. She is a Senior Researcher and a Chief Editor with the Department of Solid State Physics and Nonlinear Physics, Al-Farabi Kazakh National University. Her current research interests include signal modulation classification, signal processing, network implementation, network theory, and routing of wireless networks.
\end{IEEEbiography}

\begin{IEEEbiography}[{\includegraphics[width=1in,height=1.25in,clip,keepaspectratio]{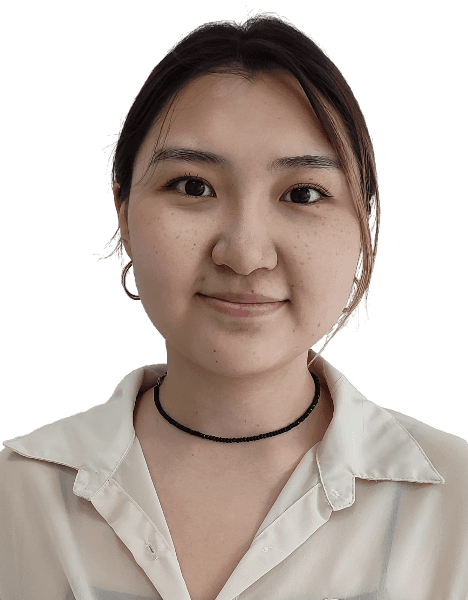}}]{S. Temesheva} received the master’s degree in radio engineering, electronics and telecommunication from Al-Farabi Kazakh National University, Almaty, Kazakhstan, in 2023.  She is currently a Researcher with the Department of Solid State Physics and Nonlinear Physics, Al-Farabi Kazakh National University. Her current research interests include signal processing, wireless networks theory, wireless communication. 
\end{IEEEbiography}

\begin{IEEEbiography}[{\includegraphics[width=1in,height=1.25in,clip,keepaspectratio]{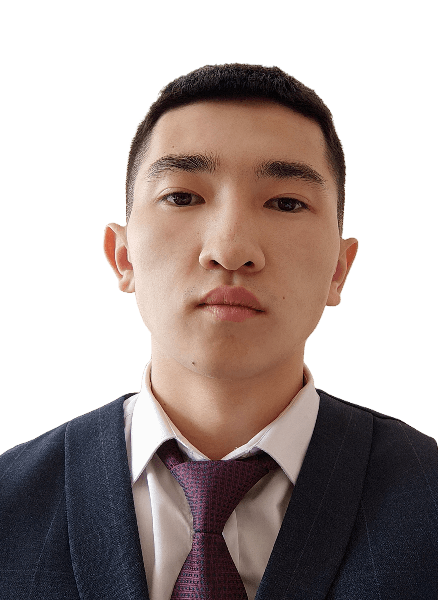}}]{A. Akhmetali} received the B.S. degree in physics from Al-Farabi Kazakh National University, Almaty, Kazakhstan, in 2024. He is a Junior Researcher with the Department of Solid State Physics and Nonlinear Physics, Al-Farabi Kazakh National University. His current research interests include signal processing, information theory, and network implementation.
\end{IEEEbiography}

\begin{IEEEbiography}[{\includegraphics[width=1in,height=1.25in,clip,keepaspectratio]{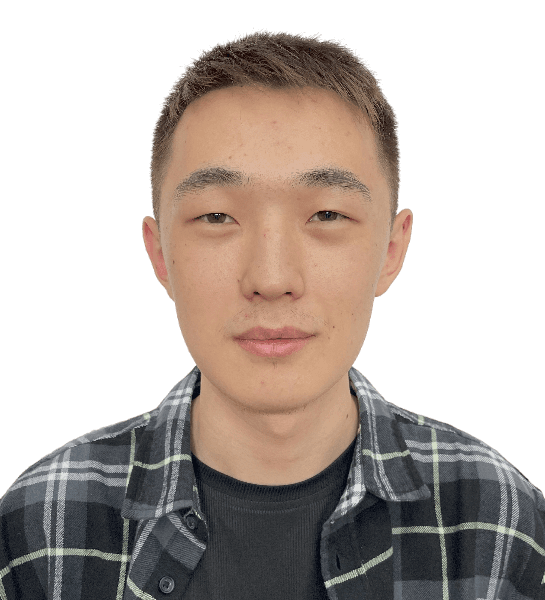}}]{M. Zaidyn} is currently pursuing the degree with Al-Farabi Kazakh National University, Almaty, Kazakhstan. He is a Junior Researcher with the Department of Solid State Physics and Nonlinear Physics, Al-Farabi Kazakh National University. His current research interests include routing algorithms, signal processing, information theory and network implementation.
\end{IEEEbiography}

\begin{IEEEbiography}[{\includegraphics[width=1in,height=1.25in,clip,keepaspectratio]{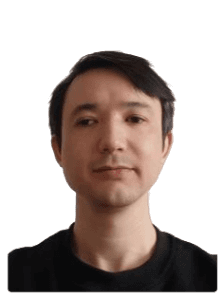}}]{T. Namazbayev} received the master’s degree in radio engineering, electronics, and telecommunication from Al-Farabi Kazakh National University, Almaty, Kazakhstan, in 2019.  He is currently a Senior Researcher at the Department of Solid State Physics and Nonlinear Physics of the Al-Farabi Kazakh National University. His current research interests include signal modulation classification, signal processing, computer modeling and network implementation.
\end{IEEEbiography}

\begin{IEEEbiography}[{\includegraphics[width=1in,height=1.25in,clip,keepaspectratio]{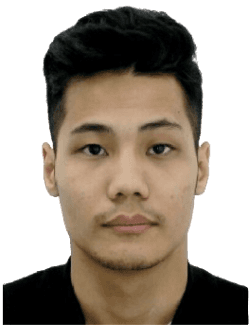}}]{A. Bolysbay} received the B.S. degree in physics from Al-Farabi Kazakh National University, Almaty, Kazakhstan, in 2024. He is a Junior Researcher with the Department of Solid State Physics and Nonlinear Physics, Al-Farabi Kazakh National University. His current research interests include signal processing, information theory, and network implementation.
\end{IEEEbiography}

\begin{IEEEbiography}[{\includegraphics[width=1in,height=1.25in,clip,keepaspectratio]{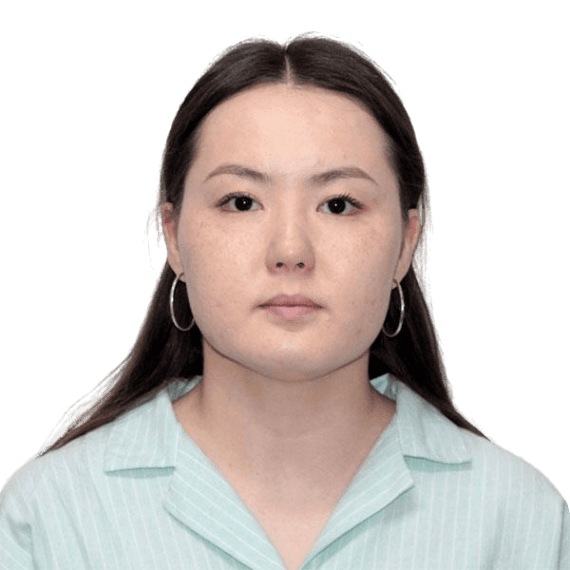}}]{A. Akniyazova} received the master’s degree in physics from Al-Farabi Kazakh National University, Almaty, Kazakhstan, in 2020.  She is currently pursuing the Ph.D degree with Al-Farabi Kazakh National University. She is a Researcher with the Department of Solid State Physics and Nonlinear Physics, Al-Farabi Kazakh National University. Her current research interests include signal processing and information theory.
\end{IEEEbiography}

\begin{IEEEbiography}[{\includegraphics[width=1in,height=1.25in,clip,keepaspectratio]{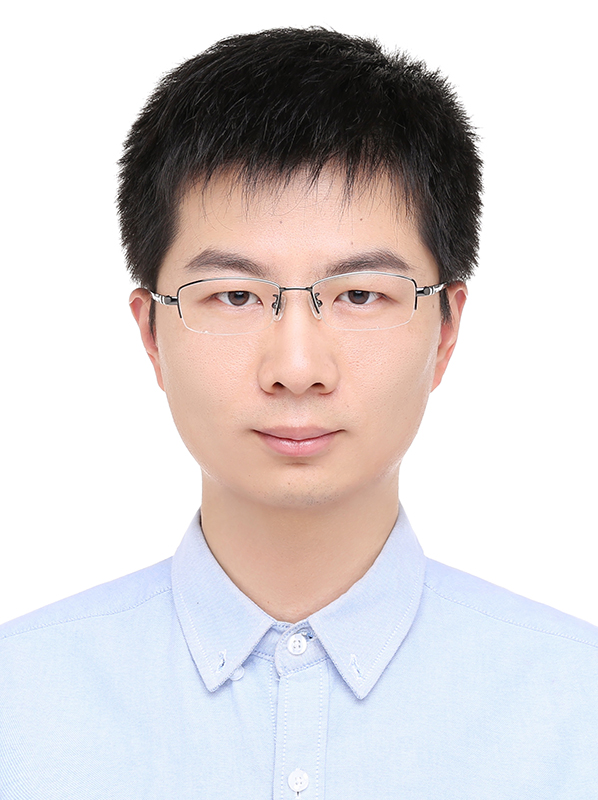}}]{Xiao Tang} (Member, IEEE) received the B.S. degree in information engineering (Elite Class named after Tsien Hsue-shen) and the Ph.D. degree in information and communication engineering from Xi’an Jiaotong University, Xi’an, China, in 2011 and 2018, respectively. He is currently with the Department of Communication Engineering, Northwestern Polytechnical University, Xi’an. His research interests include wireless communications and networking, game theory, and physical layer security.
\end{IEEEbiography}

\EOD
\end{document}